\documentclass[submission,copyright,creativecommons]{eptcs}
\usepackage{underscore} 
\usepackage{amsmath}
\usepackage{amssymb}  
\usepackage{amsthm}
\usepackage{stmaryrd}
\usepackage{multirow}

\newtheorem{theorem}{Theorem}[section]
\newtheorem{proposition}{Proposition}[section]
\newtheorem{corollary}{Corollary}[theorem]

\newtheorem{fact}{Fact}

\theoremstyle{definition}
\newtheorem{definition}{Definition}[section]

\theoremstyle{remark}
\newtheorem{remark}{Remark}
\newtheorem{example}{Example}

\newcommand{\set}[1]{\{#1\}}

\newcommand{\ar}[1]{\ll #1 \gg}
\newcommand{\br}[1]{\llbracket #1\rrbracket}

\DeclareMathOperator{\Pa}{\mathsf{Pa}}
\DeclareMathOperator{\Na}{\mathsf{Na}}
\DeclareMathOperator{\Ca}{\mathsf{Ca}}

\title{Pareto Optimality, Functional Dependence\\ and Collective Agency}
\author{Chenwei Shi
\institute{Tsinghua - Amsterdam Joint Research Centre for Logic, \\ Department of Philosophy, \\
Tsinghua University \\
Beijing, China}
\email{scw@mail.tsinghua.edu.cn}
\and Yiyan Wang
\institute{Tsinghua - Amsterdam Joint Research Centre for Logic, \\ Department of Philosophy, \\
Tsinghua University \\
Beijing, China}
\email{\quad wang-yy19@mails.tsinghua.edu.cn}
}
\def\titlerunning{Pareto Optimality, Functional Dependence and Collective Agency}
\def\authorrunning{C.Shi \& Y.Wang}
\begin{document}
\maketitle

\begin{abstract}
This paper approaches the problem of understanding collective agency from a logical and game-theoretical perspective. Instead of collective intentionality, our analysis highlights the role of Pareto optimality. To facilitate the analysis, we propose a logic of preference and functional dependence by extending the logic of functional dependence. In this logic, we can express Pareto optimality and thus reason about collective agency.

\end{abstract}

\section{Introduction}

Can a group of individual agents have agency? In philosophy of action, a widely discussed theory (cf. \cite{anscombe1957}; \cite{davidson1963}),  called ``the standard theory of agency" in \cite{sep-agency}, contends that a being has agency just in case it has the capacity to act intentionally. Sympathetic to this theory of agency, several philosophers approach the above question by analyzing a group's intentionality (cf. \cite{gilbert2006}; \cite{searle2010};  \cite{tuomela2013}; \cite{bratman2014}).  

Instead of making intentionality the starting point, our analysis focuses on ``optimality"  so as to anchor the analysis in a game-theoretical framework. In game theory, a player acts according to what she takes to be optimal for her and, normally, that depends on the other players' actions. So there is \emph{interdependence} between what is optimal for each player and thus their actions. In particular, there are two ways in which such interdependence shapes what is optimal for a player: first, by restricting what a player can do, namely actions feasible for her; second, by changing a player's preferences of those actions feasible for her. Agency of a group emerges from such interdependence when it leads the whole group to what is collectively optimal, just as an individual player's agency emerges when she makes choices she takes to be optimal. Interdependence and collective optimality are thus key to our analysis of collective agency. 

It is worth emphasizing that not modeling intentionality explicitly does not mean that the game theoretical setting is incompatible with the intention-centered understanding of agency. Some work has been done on integrating intentionality into the framework of game theory, for instance, \cite{roy2008}. 
But for the purpose of this paper, to talk about agency in game theory we take the notion of optimality as the delegate of intentionality:
\begin{quote}
An agent's \emph{intention} decides what are the \emph{optimal} choices among actions \emph{feasible} for her and the agent chooses those optimal ones if she has the capacity to exercise \emph{agency}.  
\end{quote}

To illustrate the main idea behind our analysis, consider the following example of coordination game. 

\begin{example}\label{exp:RockJazz}
Eve and Adam are a couple. There are two live clubs they frequently visited, one featuring rock and the other featuring jazz.  They plan to go out to celebrate their wedding anniversary, and before getting off work, Eve sends a message to Adam, saying ``It is our anniversary. Let's go to the live club we frequently visited after work". 
\end{example}

Depending on what Eve means by ``the live club" and how Adam understands it, there are four situations they may end up with. As Table \ref{tab:RoJ} shows, they both prefer going to the same live club to going to different ones and prefer going together to the Jazz club to going to the Rock club. It is clear that what is optimal for Eve is dependent on Adam's choice, vice versa.\footnote{This is the second type of interdependence. For an example of the first type of interdependence, consider the scenario where Eve and Adam are having dinner in a restaurant and need to order their desserts. They have to order the same dessert, so Eve says to Adam `` You decide". In this extreme case, Adam's choice decides Eve's choice and thus what is optimal for her.}

\begin{table}
\centering
\begin{tabular}{llll}
                                             & \multicolumn{1}{c}{}       & \multicolumn{2}{c}{Adam}                              \\ \cline{3-4} 
                                             & \multicolumn{1}{l|}{}      & \multicolumn{1}{l|}{Jazz} & \multicolumn{1}{l|}{Rock} \\ \cline{2-4} 
\multicolumn{1}{c|}{\multirow{2}{*}{Eve}} & \multicolumn{1}{l|}{Jazz} & \multicolumn{1}{l|}{4,4}   & \multicolumn{1}{l|}{0,0}   \\ \cline{2-4} 
\multicolumn{1}{c|}{}                        & \multicolumn{1}{l|}{Rock} & \multicolumn{1}{l|}{0,0}   & \multicolumn{1}{l|}{1,1}   \\ \cline{2-4} 
\end{tabular} 
 
\caption{Rock or Jazz?}\label{tab:RoJ}
\end{table}

Suppose that they both went to the rock club. In this case, intuitively, going together to the jazz club should be collectively optimal for the couple because going together to the jazz club is better for both of them than all the other three states. According to the above understanding of agency, if the couple fails to reach the collectively optimal state, then they fail to have collective agency.

To make our analysis of collective agency precise, we propose a logic of preference and functional dependence. In this logic, we can express and reason about interdependence and thus define Nash equilibrium and Pareto optimality. These central notions in game theory pertain to our analysis of collective optimality. Our logic is a natural extension and combination of the logic of functional dependence in \cite{johan2021} and the basic modal preference logic in \cite[Section 2]{vanBenthem2007}. It can also be seen as a further development of the modal logical approach to analyzing strategic games proposed in \cite{johan2011}, in the sense of not having the independence assumption and thus making dependence between players' actions explicit.\footnote{The logic in \cite{johan2011} takes knowledge into account, which is left out in our logic for simplicity.}

\paragraph{Contributions of this paper} Technically, we develop a logic of preference and functional dependence which can be used to characterize strategic games. Moreover, we present a sound and complete Hilbert-style axiom system for the logic. Conceptually, we show how collective agency can emerge from interdependence between what is optimal for each group member. By emphasizing the role of optimality rather than intentionality, we bring a game theoretical perspective to the problem. Finally,  we present and evaluate several candidate definitions of collective agency and propose a formal definition.


\paragraph{Structure of the paper} We first introduce a logic of preference and functional dependence, including its Hilbert-style axioms system in Section \ref{sec:LPFD} and then demonstrate how Nash equilibrium and weak/strong Pareto optimality can be expressed in the logic in Section \ref{sec: POandNEinLPFD}. Then in Section \ref{sec:CAandPO}, we analyze collective agency in the logic. Conclusions and indications of further work can be found in Section \ref{sec:CandFW}.

\section{The Logic of Preference and Functional Dependence (LPFD)}\label{sec:LPFD}

We first introduce the syntax and semantics of the logic of functional dependence (LFD) in \cite{johan2021} and then extend it with components for dealing with preference. For the purpose of this paper, we customize LFD to suit the narrative of strategic games. No substantial changes are made to the original setting of LFD until we extend it with preference orders (Definition \ref{def:PDmodel}) and coin new operators (Definition \ref{def:PDlanguage}). 

LFD starts with a set of variables $V$ and a domain of objects $O$. We take $V$ as the set of players in a game and $O$ as the set of actions each player can perform in the game. Then a set of admissible assignments of actions to players $A\subseteq O^V$ can be collected to represent possible strategy profiles of the game.  In addition, a relational vocabulary $(Pred,ar)$ is given to describe these possible strategy profiles, where $Pred$ is a set of predicate symbols and $ar: Pred\rightarrow N$ is an arity map, associating to each predicate $P\in Pred$ a natural number $ar(P)$.

\begin{definition}[Dependence models]\label{def:Dmodel}
A dependence model $\mathbf{M}$ is a pair $\mathbf{M} = (M,A)$ of a (relational) first order logic $M = (O,I)$ with a domain of actions $O$ and an interpretation map $I$, together with a set of strategy profiles $A\subseteq O^V$. The interpretation map $I$ assigns to each predicate $P\in Pred$ a subset of $O^{\ ar(P)}$.
\end{definition} 
In a dependence model, when $A\neq O^V$, some strategy profiles are missing. This gives rise to dependence between players' actions. Suppose a strategy profile $s$ for two players $x$ and $y$ is not in $A$. Then $x$ and $y$ cannot act according to $s$ simultaneously. In some sense this form of dependence is weak because it does not differentiate between different types of dependence, for example, correlation and causation. However, the other side of the same coin is its generality which is helpful for capturing some common properties of different types of dependence. For further explanation of how and what kinds of dependence can be captured in a dependence model, we refer readers to \cite{johan2021}. 

To capture functional dependence, LFD uses two operators $\mathbb{D}_X \varphi$ and $D_X y$ in its language. 
\begin{definition}
Given a vocabulary $(V,Pred,ar)$, the language LFD $\mathcal{L}$ is given by
$$\varphi :: = P\mathbf{x} \mid \neg\varphi \mid \varphi\wedge\varphi \mid \mathbb{D}_X \varphi \mid D_X y$$
where $P\in Pred$, $\mathbf{x} = (x_1,\ldots, x_n)$ is a finite string of players of length $n = ar(P)$ and $X\subseteq V$ is a finite set of players and $y\in V$ is a player. 
\end{definition}
$\mathbb{D}_X \varphi$ says that whenever the players in $X$ takes their current actions, $\varphi$ is the case;  $D_X y$ says that whenever the players in $X$ take their current actions, $y$ also takes its current action.
\begin{definition}
Truth of a formula $\varphi\in \mathcal{L}$ in a dependence model $\mathbf{M} = (M,A)$ at a strategy profile $s\in A$ is defined as follows:
\begin{center}
\begin{tabular}{lll}
$\mathbf{M},s\models P\mathbf{x}$ & iff & $s(\mathbf{x})\in I(P)$\\
$\mathbf{M},s\models \neg \varphi$ & iff & $\mathbf{M},s\not\models \varphi$\\
$\mathbf{M},s\models \varphi\wedge \psi$ & iff & $\mathbf{M},s\models \varphi$ and $\mathbf{M},s\models \psi$\\
$\mathbf{M},s\models \mathbb{D}_X\varphi$ & iff & $\mathbf{M}, t\models \varphi$ holds for all $t\in A$ with $s =_X t$ \\
$\mathbf{M},s\models D_X y$ & iff & $ s=_y t$ for all $t\in A$ with $s =_X t$\\
\end{tabular}
\end{center}
where $s =_X t$ if and only if for each $x\in X$, the action of $x$ in $s$ is the same as her action in $t$ and $s=_y t$ is the abbreviation for $s=_{\set{y}} t$. Note that $=_X$ is an equivalence relation on $A$ and $s=_\emptyset t$ holds for all $s,t\in A$.
\end{definition}

Next, we extend LFD to LPFD. 
\begin{definition}[Preference-dependence models]\label{def:PDmodel}
A preference-dependence (PD) model $\mathbb{M}$ is a pair $\mathbb{M} = (\mathbf{M},\set{\preceq_x}_{x\in V})$ of a dependence model $\mathbf{M}$ and a reflexive and transitive order on the set of strategy profiles $A\in \mathbf{M}$ for each player $x\in V$. 

We will write $s\preceq_X t$ for ``$s\preceq_x t$ for all $x\in X$" and $s\prec_X t$ for ``$s\prec_x t$ for all $x\in X$". Especially, we will write $s\simeq_X t$ for ``$s\preceq_X t$ and $t\preceq_X s$" and $s\precnsim_X t$ for ``$s\preceq_X t$ and $t\npreceq_X s$". Note that $\prec_x$ is the same as $\precnsim_{\set{x}}$, but $\prec_X$ is different from $\precnsim_X$.
\end{definition} 
\begin{definition}(Syntax)\label{def:PDlanguage}
Given a vocabulary $(V, Pred, ar)$, the language of LPFD $\mathcal{L}_P$ is generated by the following grammar:
 $$\varphi ::= P\textbf{x}\ |\ \neg\varphi\ |\ \varphi\wedge\psi\ |\ \br{=_{X},\preceq_{X'},\prec_{X''}} \varphi\ |\ [=_{X},\preceq_{X'},\prec_{X''}]  y $$
The duality of $\br{=_{X},\preceq_{X'},\prec_{X''}} \varphi$ is written as $\ar{=_{X},\preceq_{X'},\prec_{X''}} \varphi$ and $\bigwedge_{y\in Y} [=_{X},\preceq_{X'},\prec_{X''}] y$ is abbreviated to $[=_{X},\preceq_{X'},\prec_{X''}]  Y$.
\end{definition}
\begin{definition}
Truth of a formula $\varphi\in \mathcal{L}$ in a PD model $\mathbb{M} = (\mathbf{M},\set{\preceq_x}_{x\in V})$ at a strategy profile $s\in A$ is defined as follows (with the atomic and Boolean cases defined as in LFD):
\begin{center}
\begin{tabular}{lll}
$\mathbb{M}, s \models \br{=_{X},\preceq_{X'},\prec_{X''}} \varphi$ & iff & for all $t\in A$ satisfying $s =_{X} t$, $s \preceq_{X'} t$ and $s\prec_{X''} t$, $t \models \varphi$; \\
$\mathbb{M}, s \models [=_{X},\preceq_{X'},\prec_{X''}] y$ & iff & for all $t\in A$ satisfying $s =_{X} t$, $s \preceq_{X'} t$ and $s\prec_{X''} t$, $s =_y t$,
\end{tabular}
\end{center}
\end{definition}
Note that we can define $\mathbb{D}_X\varphi$ and $D_X y$ as $\br{=_{X},\preceq_\emptyset,\prec_\emptyset} \varphi$ and $[=_{X},\preceq_\emptyset,\prec_\emptyset] y$  in $\mathcal{L}_P$ respectively. $\br{=_\emptyset,\preceq_x,\prec_\emptyset} \varphi$ and $\br{=_\emptyset,\prec_\emptyset,\prec_x} \varphi$ are standard modal operators defined on $\preceq_x$ and $\prec_x$ respectively. Thus $\br{=_X,\preceq_{X'},\prec_{X''}} \varphi$ is in fact a standard modal operator defined on the intersection of the relations $=_X$, $\preceq_{X'}$ and $\prec_{X''}$. 

We have mentioned that there are two types of interdependence between players in a game. In LPFD, the first type, which comes from restricting what a player can do, can be captured by the operators $\mathbb{D}_X$ and $D_X$; the second type is captured by $\br{=_{X},\preceq_{X'},\prec_{X''}}$ and $[=_{X},\preceq_{X'},\prec_{X''}]$, because it concerns how players' preferences change. We formalize Example \ref{exp:RockJazz} to illustrate how the second type of interdependence in a game is captured in LPFD.

\begin{example}\label{exp:Frockjazz}
To get a PD model, take $O = \set{R,J}$ where $R$ means going to the rock club and $J$ means going to the jazz club; $A = \set{s_1=RR,s_2=RJ,s_3=JR,s_4=JJ}$ where $RJ$ means that Eve goes to the rock club and Adam goes to the jazz club; for Eve, the preference order $\preceq_E$ is given by $s_2 \simeq_E s_3\prec_E s_1\prec_E s_4$, which is the same as $\preceq_A$ for Adam.  $I$ does not play a role in our example and can be specified arbitrarily. 
\end{example}

In the model, $s_1\models \br{=_E,\preceq_\emptyset,\prec_A}\bot \wedge \br{=_A,\preceq_\emptyset,\prec_E}\bot$; $s_4\models \br{=_E,\preceq_\emptyset,\prec_A}\bot \wedge \br{=_A,\preceq_\emptyset,\prec_E}\bot$; $s_4\models \br{=_\emptyset,\preceq_\emptyset,\prec_{\set{E,A}}}\bot$. 
The first and second facts say that given one of the couple goes to one of the two live clubs, there is no better choice for the other one than going to the same live club. So what is optimal for one of the couple depends on what the other one does. The third facts says that there is no other state which is better for both Eve and Adam than going together to the jazz club.

\paragraph{Axiom System}\quad The Hilbert-style axiom system of \textbf{LPFD} in Table \ref{tab:axiomSystem} can be seen as a combination of the axiom system of LFD in \cite{johan2021}, and the axioms characterizing the interaction between the standard modal operators defined on $\preceq_x$ and $\prec_x$ in \cite{vanBenthem2007} with the necessary adaptation to the more general operators $\br{=_{X},\preceq_{X'},\prec_{X''}}$ and $[=_{X},\preceq_{X'},\prec_{X''}]$.  The function of $Free(\varphi)$ used in the axiom \textbf{(II)(c)} is defined recursively as follows: $Free(Px_1\ldots x_n) = \set{x_1,\ldots,x_n}$, $Free(\neg\varphi) = Free(\varphi)$, $Free(\varphi\wedge \psi) = Free(\varphi)\cup Free(\psi)$, $Free(\br{=_{X},\preceq_{X'},\prec_{X''}}\varphi) = X$ and $Free([=_{X},\preceq_{X'},\prec_{X''}] y) = X$ 
\newcommand{\tabincell}[2]{\begin{tabular}{@{}#1@{}}#2\end{tabular}}
\begin{table}[!t]
	\centering
	\scriptsize

	\begin{tabular}{ll}
		\\[-2mm]
		\hline
		\hline\\[-2mm]
		
		\vspace{3mm}\\[-3mm]
\textbf{(I)} & Axioms and rules for classical proposition logic\\

		\hline	\\[-2mm]
		
\textbf{(II)} & Axioms and rules for $\br{=_{X},\preceq_{X'},\prec_{X''}}$\\
\textbf{(a)} & from $\varphi$ infer $\br{=_{X},\preceq_{X'},\prec_{X''}} \varphi$\\
\textbf{(b)} & $\br{=_{X},\preceq_{X'},\prec_{X''}} (\varphi\rightarrow \psi)\wedge (\br{=_{X},\preceq_{X'},\prec_{X''}} \varphi\rightarrow \br{=_{X},\preceq_{X'},\prec_{X''}} \psi$)\\
\textbf{(c1)} & $\varphi\rightarrow \br{=_{X},\preceq_\emptyset,\prec_\emptyset} \varphi$ provided that $Free(\varphi)\subseteq X$\\
\textbf{(c2)} & $\br{=_{X},\preceq_{X'},\prec_{X''}}\varphi\rightarrow \br{=_{X},\preceq_{X'},\prec_{X''}}\br{=_{X},\preceq_{X'},\prec_{X''}}\varphi$\\
\textbf{(d)} & $\br{=_{X},\preceq_{X'},\prec_\emptyset}\varphi\rightarrow \varphi$\\
\textbf{(e)} & $\br{=_{X},\preceq_{X'},\prec_{X''}}\varphi\rightarrow \br{=_{Y},\preceq_{Y'},\prec_{Y''}}\varphi$ provided that $X\subseteq Y$, $X'\subseteq Y'$ and $X''\subseteq Y''$\\[2mm]
& Axioms characterizing the relations between $\preceq_x$ and $\prec_x$\\
\textbf{(f)} & $\ar{=_{X},\preceq_{X'},\prec_{X''}}\varphi\rightarrow\  \ar{=_{X},\preceq_{X'\cup X''},\prec_{X''}}\varphi$\\
\textbf{(g)} & $\ar{=_{X},\preceq_{X'},\prec_{X''}}\ar{=_{X},\preceq_{X'- Y},\prec_{X''\cup Y}}\varphi\rightarrow\ \ar{=_{X},\preceq_{X'},\prec_{X''\cup Y}}\varphi$ provided that $Y\subseteq X'$\\
\textbf{(h)} & $\ar{=_{X},\preceq_{X'},\prec_{X''}}\ar{=_{X},\preceq_{X'\cup Y},\prec_{X''- Y}}\varphi\rightarrow\ \ar{=_{X},\preceq_{X'\cup Y},\prec_{X''}}\varphi$ provided that $Y\subseteq X''$\\
\textbf{(j)} &  $(\varphi\wedge \ar{=_{X},\preceq_{X'},\prec_{X''}}\psi)\rightarrow ((\bigvee_{x\in X'} \ar{=_{X},\preceq_{X'},\prec_{X''\cup\set{x}}}\psi)\ \vee \ar{=_{X},\preceq_{X'},\prec_{X''}}(\psi\wedge \ar{=_{X},\preceq_{X'},\prec_\emptyset}\varphi))$\\
	    
       \hline	\\[-2mm]
       
\textbf{(III)} & Axioms and rules for $[=_{X},\preceq_{X'},\prec_{X''}]$\\
\textbf{(a)} & $[=_{X},\preceq_{X'},\prec_{X''}]x$ provided that $x\in X$\\
\textbf{(b)} & $[=_{X},\preceq_{X'},\prec_{X''}]Y\wedge [=_{Y},\preceq_{X'},\prec_{X''}]Z\rightarrow [=_{X},\preceq_{X'},\prec_{X''}]Z$\\
    \hline	\\[-2mm]
       
\textbf{(IV)} & Axioms and rules for $\br{\ }-[\ ]$ interaction\\
\textbf{(a)} & $[=_{X},\preceq_{X'},\prec_{X''}]Y\wedge \br{=_{Y},\preceq_{X'},\prec_{X''}}\varphi\rightarrow \br{=_{X},\preceq_{X'},\prec_{X''}}\varphi$\\[2mm]
		\hline
		\hline\\
	\end{tabular}
	\caption{The Hilbert-style proof system  {\bf LPFD}}\label{tab:axiomSystem}
\end{table}

\begin{proposition}\label{prop:soundness}
The axiom system \textbf{LPFD} is sound for PD models.
\end{proposition}
\begin{proof}
We take the axioms \textbf{(IV)(a)} and \textbf{(II)(j)} as two examples,  showing their validity and how \textbf{LPFD} generalizes the axioms in \textbf{LFD} and the axioms characterizing the interaction between $\preceq_x$ and $\prec_x$ in \cite{vanBenthem2007}.

The axiom \textbf{(IV)(a)} generalizes the axiom \textbf{Transfer}: $(D_XY\wedge \mathbb{D}_Y\varphi)\rightarrow \mathbb{D}_X\varphi$ in \textbf{LFD}. Its validity follows from the fact that given its  antecedent is satisfied on an arbitrary strategy profile $s\in A$ in an arbitrary PD model, $\set{t\in A\mid s=_X t, s\preceq_{X'} t,  s\prec_{X''} t}\subseteq \set{t\in A\mid s=_Y t, s\preceq_{X'} t,  s\prec_{X''} t}\subseteq \set{t\in A\mid t\models \varphi}$.

The axiom \textbf{(II)(j)} generalizes the axiom $\mathbf{Int}_2$: $\varphi\wedge \Diamond^{\leq}\psi\rightarrow (\Diamond^<\psi\vee\Diamond^{\leq}(\psi\wedge \Diamond^{\leq}\varphi)$ in \cite{vanBenthem2007}. Note that the logic in \cite{vanBenthem2007} deals with only a single agent and thus no subscripts for different agents are needed for the modal operators in $\mathbf{Int}_2$. Our logic deals with not only preference orders for multiple agents but also the intersection of these preference orders. Semantically, the axiom $\mathbf{Int}_2$ means that if on the current state $s$, $\varphi$ is satisfied and in a state $t$ at least as good as the current state $\psi$ is satisfied, then either $t$ is strictly better than $s$ (so $\Diamond^<\psi$ is satisfied) or $t$ and $s$ are equally good  (so $\Diamond^{\leq}(\psi\wedge \Diamond^{\leq}\varphi)$ is satisfied). To see how the axiom \textbf{(II)(j)} generalizes the axiom $\mathbf{Int}_2$, it is suffice to realize that when there is a state in $\set{t\in A\mid s=_X t, s\preceq_{X'} t,  s\prec_{X''} t}$ satisfying $\varphi$ as the truth of $\ar{=_{X},\preceq_{X'},\prec_{X''}}\psi$ on $s$ requires, it is either in $\bigcup_{x\in X'}\set{t\in A\mid s=_X t, s\preceq_{X'} t,  s\prec_{X''\cup\set{x}} t}$ or in $\set{t\in A\mid s=_X t, s\simeq_{X'} t,s\prec_{X''}t}$ (thus in $\set{t\in A\mid s=_X t, s\simeq_{X'} t}$).
\end{proof}

While the soundness of \textbf{LPFD} is not hard to prove, the proof of the completeness of \textbf{LPFD} is not trivial. Nevertheless, there is a way to follow, as established in Appendix A of \cite{johan2021}, by making use of several standard techniques for transforming models.
\begin{theorem}
The axiom system \textbf{LPFD} is strongly complete for PD models.
\end{theorem}
\begin{proof}
Instead of elaborating on the details of the proof here, we will provide them in the full paper. We only note that the proof hinges on two pivotal steps:
first, a proper adaption of both the standard relational model \cite[Definition 3.13]{johan2021} and the relational model \cite[Definition A.1]{johan2021}; second, a proper adaption of \textit{unraveling}, the technique of transforming models.
\end{proof}


\begin{remark}[\textbf{Open question}]\label{rmk:openQ}
One caveat to our above ``no difficulties" remark: when $\preceq_x$ is required to be total, which is a very common assumption in the literature on game theory, the technique of unraveling does not work anymore, because unraveling breaks the totalness of the relation $\preceq_x$.  
Therefore, the existence of a sound and complete axiom system for LPFD with respect to the class of total PD models (which require $\preceq_x$ for each $x$ to be total) is still an open question.
\end{remark}

\section{Pareto Optimality and Nash Equilibrium Expressed in LPFD}\label{sec: POandNEinLPFD}
Having laid out the basics of LPFD, in this and next sections, we turn to questions concerning expressing and reasoning about Pareto optimality, Nash equilibrium, and related issues in LPFD. One important assumption we will adopt is that the group of players $V$ has to be finite. In LPFD, there is no such a restriction on $V$. However, it is worth noting that in the language of LPFD all subscripts in the two operators need to be finite. So to express something like $\br{=_{-X},\preceq_\emptyset,\prec_X} \varphi$ in LPFD where $-X := V-X$, which is frequently referred to in game theory,  we have to ensure that $X$ and $-X$ are both finite.

We start with recalling what Nash equilibrium and weak/strong Pareto optimality mean.
\begin{definition}\label{def:NwsP}
In a PD model $\mathbb{M}$, given that the players in $-X$ have acted according to the strategy profile $s\in A$,
\begin{itemize}
\item  $s$ is a \textbf{Nash equilibrium} for $X\subseteq V$ if for all $x\in X$ there is no $t=_{-x} s$ such that $s\prec_x t$;
\item  $s$ is \textbf{(strongly) Pareto optimal} for $X\subseteq V$ if there is no $t=_{-X} s$ such that (a) for all $x\in X$, $s\preceq_x t$ and (b) there is one $x\in X$ such that $s\prec_x t$;
\item  $s$ is \textbf{(weakly) Pareto optimal} for $X\subseteq V$  if there is no $t=_{-X} s$ such that for all $x\in X$, $s\prec_x t$.
\end{itemize}
\end{definition}
Note that such a way of defining the notions of Nash equilibrium, weak and strong Pareto optimality in a PD model applies to all subgroups of $V$ rather than only the whole group of players $V$. In Example \ref{exp:RockJazz}, going together to the rock club and going together to the jazz club are two Nash equilibria and only going together to the jazz club is Pareto optimal, both weakly and strongly.

It is relatively easy to get how Nash equilibrium and weak Pareto optimality can be expressed in LPFD, as the following fact shows.
\begin{fact}\label{fact:WPandNA}
In a PD model $\mathbb{M}$,
\begin{itemize}
\item $s$ is a Nash equilibrium for $X\subseteq V$ given that the players in $-X$ have acted according to $s$ \textbf{iff}
$\mathbb{M}, s\models \bigwedge_{x\in X}\br{=_{-X},\preceq_\emptyset,\prec_x}\bot$;
\item $s$ is weakly Pareto optimal for $X\subseteq V$ given that the players in $-X$ have acted according to $s$ \textbf{iff}
$\mathbb{M}, s\models \br{=_{-X},\preceq_\emptyset,\prec_{X}}\bot$.
\end{itemize}
\end{fact}
In the case of weak Pareto optimality, because the truth condition of the operator  $\br{=_{-X},\preceq_\emptyset,\prec_X}$ depends on what formulas are satisfied on all elements in the set $\set{t\in A\mid s=_{-X} t, s\prec_X t}$, if it is an empty set and thus $\bot$ can be vacuously satisfied on all elements in it, then $s$ is weakly Pareto optimal for $X$.  In Example \ref{exp:Frockjazz}, the first and second facts say that $s_1$ and $s_4$ are two Nash equilibria while the third one says that $s_4$ is weakly Pareto optimal. 

To express strong Pareto optimality in LPFD, we need to express that the following model theoretical fact, namely the set $\set{t\in A\mid s=_{-X} t, s\precnsim_X t} = \bigcup_{x\in X}\set{t\in A\mid s=_{-X} t, s\preceq_{X-\set{x}} t,s \prec_{x} t}$ is empty. Since $s\models \br{=_{-X},\preceq_{X-\set{x}}, \prec_{x}}\bot$ if and only if $\set{t\in A\mid s=_{-X} t, s\preceq_{X-\set{x}},\prec_{x}} = \emptyset$, we can define strong Pareto optimality as follows. 
\begin{fact}\label{fact:SP}
In a PD model $\mathbb{M}$,
$s$ is strongly Pareto optimal for $X\subseteq V$ given that the players in $-X$ have acted according to $s$ \textbf{if and only if} $\mathbb{M}, s\models \bigwedge_{x\in X}\br{=_{-X},\preceq_{X-\set{x}}, \prec_{x}}\bot$.
\end{fact}

\noindent To facilitate our discussion, we define strong Pareto optimality in LPFD as
$$\Pa X := \bigwedge_{x\in X}\br{=_{-X},\preceq_{X-\set{x}}, \prec_{x}}\bot\enspace .$$
An easy observation is that we can define Nash equilibrium in terms of strong Pareto optimality as $$\Na X := \bigwedge_{x\in X} \Pa \set{x}\enspace .$$
This fact will also be helpful in the next section for our discussion about collective agency.
 
\section{Collective Agency and Pareto Optimality}\label{sec:CAandPO}

In this section, we analyze several different candidate definitions of collective agency and then propose a formal definition in LPFD.

\paragraph{Collective Optimality}  We have mentioned in the introduction that an agent acts according to what her intention prescribes as optimal. Analogously, collective agency should drive the group towards a collectively optimal state:
\begin{quote}
A group demonstrates collective agency if it ends up with the \emph{collectively optimal} strategy profiles among those feasible for it.
\end{quote}
In the case of individual agency, what is optimal is self-evident in a game model because each player's preference order is given in the model. However, in the case of collective agency, while a group's collective actions are given by the strategy profiles, it is not so clear which strategy profiles are collectively optimal. There is no preference order for the group given in the model. This is advantageous to a further analysis, because it leaves room for different ways of defining collective agency, depending on how collective optimality is defined.\footnote{Social choice theory provides another aproach to the question of defining collective optimality. The work in \cite{list2011} makes use of social choice theory to help understand group agency.} 

Notwithstanding different possible definitions of collective optimality, we contend that a reasonable one should imply Pareto optimality. Suppose that a group ends up with a strategy profile $s$ which is not Pareto optimal. Then there is another strategy profile $t$ such that every player in the group takes $t$ to be at least as good as $s$ and at least one player prefers $t$ to $s$. In this situation, everyone in the group is willing to act according to $t$ but do not coordinate with each other to act according to $t$. It seems reasonable to maintain that a group whose members do not coordinate to reach a better state does not have any collective agency.

\paragraph{Collective Agency Based on Pareto Optimality} Does Pareto optimality suffice for collective agency? Suppose it suffices.  Then we can interpret $s\models \Pa X$ as ``X has collective agency at $s$". Specifically, if $X = \set{x}$ in $\Pa X$, it says that a single player $x$ has agency at $s$. Now comes an important question. If $X$ has collective agency at $s$, should $x\in X$ also have agency at $s$? Intuitively, yes. It is paradoxical to have a group with collective agency while its members have no individual agency. However, game theory makes it clear that such a paradox exists if we take Pareto optimality as collective agency. The classical prisoners' dilemma (Table \ref{tab:PD1}) provides us with such a counterexample. We quote the narrative of this dilemma in \cite{sepprisonerdilemma}:
\begin{quotation}
Tanya and Cinque have been arrested for robbing the Hibernia Savings Bank and placed in separate isolation cells. Both care much more about their personal freedom than about the welfare of their accomplice. A clever prosecutor makes the following offer to each: ``You may choose to confess or remain silent. If you confess and your accomplice remains silent I will drop all charges against you and use your testimony to ensure that your accomplice does serious time. Likewise, if your accomplice confesses while you remain silent, they will go free while you do the time. If you both confess I get two convictions, but I'll see to it that you both get early parole. If you both remain silent, I'll have to settle for token sentences on firearms possession charges. If you wish to confess, you must leave a note with the jailer before my return tomorrow morning."
\end{quotation} 
\begin{table}[h]
\centering
\begin{tabular}{llll}
                                             & \multicolumn{1}{c}{}       & \multicolumn{2}{c}{Prisoner2}                              \\ \cline{3-4} 
                                             & \multicolumn{1}{l|}{}      & \multicolumn{1}{l|}{cooperate} & \multicolumn{1}{l|}{confess} \\ \cline{2-4} 
\multicolumn{1}{c|}{\multirow{2}{*}{Prisoner1}} & \multicolumn{1}{l|}{cooperate} & \multicolumn{1}{l|}{2,2}   & \multicolumn{1}{l|}{0,4}   \\ \cline{2-4} 
\multicolumn{1}{c|}{}                        & \multicolumn{1}{l|}{confess} & \multicolumn{1}{l|}{4,0}   & \multicolumn{1}{l|}{1,1}   \\ \cline{2-4} 
\end{tabular} 
\caption{Prisoners' Dilemma I}\label{tab:PD1}
\end{table}
In such a model $(\text{coop,coop})\models \Pa\set{1,2}$ but neither $(\text{coop,coop})\models \Pa\set{1}$ nor $(\text{coop,coop})\models \Pa\set{2}$.

To remedy the problem of having collective agency of $X$ while losing its members' individual agency, we can strengthen the definition of collective agency as follows: $$\Ca_1 X := \Pa X \wedge \bigwedge_{x\in X} \Pa \set{x}\enspace .$$
Under this new definition, $(\text{coop,coop})\not\models \Ca_1 X$, namely $X$ has no collective agency at $(\text{coop,coop})$. Recall that we have defined Nash equilibrium $\Na X$ as $\bigwedge_{x\in X} \Pa \set{x}$. So $\Ca_1 X$ actually requires a group to reach a Pareto optimal Nash equilibrium. 

Still, as a definition of collective agency, $\Ca_1 X$ seems controversial. Consider again the prisoners' dilemma as described in the above quotation. This time we take the prosecutor into account and model him as the third player in the game as in Table \ref{tab:PD2}.
\begin{table}[h]
\centering
\begin{tabular}{llll}
                                             & \multicolumn{1}{c}{}       & \multicolumn{2}{c}{Prisoner2}                              \\ \cline{3-4} 
                                             & \multicolumn{1}{l|}{}      & \multicolumn{1}{l|}{cooperate} & \multicolumn{1}{l|}{confess} \\ \cline{2-4} 
\multicolumn{1}{c|}{\multirow{2}{*}{Prisoner1}} & \multicolumn{1}{l|}{cooperate} & \multicolumn{1}{l|}{2,2,0}   & \multicolumn{1}{l|}{0,4,1}   \\ \cline{2-4} 
\multicolumn{1}{c|}{}                        & \multicolumn{1}{l|}{confess} & \multicolumn{1}{l|}{4,0,1}   & \multicolumn{1}{l|}{1,1,4}   \\ \cline{2-4} 
\end{tabular} 
\caption{Prisoners' Dilemma II}\label{tab:PD2}
\end{table}
Because the prosecutor does not have choices to make but has his preference  on the different strategy profiles of the two prisoners, we only need to add numbers representing his preference order. In this new model $(\text{conf,conf})\models \Pa\set{1,2,3}\wedge \Na\set{1,2,3}$. According to our revised definition, the prosecutor and the two prisoners as a group have collective agency. How come that by adding a player whose interest is almost the opposite of that of the original players the whole group are endowed with collective agency? Different answers can be given, depending on different intuitions.

For those who share the intuition that $\set{1,2,3}$ does \emph{not} have collective agency at $(\text{conf,conf})$, $\Ca_1 X$ is not a satisfactory definition of collective agency. They may propose a stronger definition as follows to resolve the problem: $$\Ca_2 X := \bigwedge_{X'\subseteq X} \Pa X'\enspace .$$
This definition requires that all subgroups of a group with collective agency are in their Pareto optimal states. 

For those who think the group $\set{1,2,3}$ can have collective agency at $(\text{conf,conf})$, the following analogy may serve as an explanation. If the whole group $\set{1,2,3}$ is taken as a company where the prosecutor is the boss and the prisoners are his workers. The prosecutor's offer is similar in spirit to what is offered by a boss to his workers. The collective agency, if the whole company has any, is mainly executed by the boss and often disadvantageous to the workers. This is why the workers need their union. This explanation seems to defend $\Ca_1 X$. 

We agree with the explanation and thus share the intuition that the group $\set{1,2,3}$ can have collective agency at $(\text{conf,conf})$. However, we are also sympathetic to the feeling of uneasiness behind the other intuition. $\Ca_1 X$ seems to impose few requirements on how the subgroups of $X$ should behave. In the extreme case, it is possible that $\Ca_1 X$ holds while $\Pa X'$ does not hold for any non-singleton set $X'\subsetneq X$.\footnote{It is not hard to get such a model by adjusting the numbers in Table \ref{tab:PD2}.} But $\Ca_2 X$ seems to be an overreaction to the problem of $\Ca_1 X$. 

\paragraph{Improving on $\Ca_1 X$ and $\Ca_2 X$}\quad  Recall that $\Pa X := \bigwedge_{x\in X}\br{=_{-X},\preceq_{X-\set{x}}, \prec_{x}}\bot$. To generalize it, we define $$\Pa_YX := \bigwedge_{x\in X}\br{=_{Y},\preceq_{X-\set{x}}, \prec_{x}}\bot\enspace .$$
So $\Pa X$ can be seen as an abbreviation of $\Pa_{-X}X$. Note that the truth condition of $\Pa_YX$ is a direct generalization of the definition of strong Pareto optimality in Definition \ref{def:NwsP} where $-X$ is replaced by $Y$.

Let $\mathfrak{X}(X) := 2^X-\set{X}$. We say that $\mathfrak{C}(X)\subseteq \mathfrak{X}(X)$ is a \emph{cover} of $X$ if $\bigcup_{X'\in \mathfrak{C}(X)} X' = X$. We define collective agency in LPFD as follows. 
\begin{definition}
We say that a group $X$ has collective agency at $s$, denoted by $\Ca X$, if there is a cover $\mathfrak{C}(X)$ of $X$ such that $s\models \Na X\wedge \bigwedge_{X'\in \mathfrak{C}(X)}\Pa_{-X}X'$.
\end{definition}
The definition says that $X$ has collective agency at its current state $s$ if (1) $s$ is a Nash equilibrium  for $X$ and (2) there is a cover of $X$ such that for all subgroups in the cover, $s$ is strongly Pareto optimal \emph{with respect to all possible strategy profiles after fixing the actions of players in $-X$}.
To see the subtlety involved in the big conjunction part of the definition, let's compare it with the formula: $\bigwedge_{X'\in \mathfrak{C}(X)}\Pa_{-X'}X'$. The only difference lies in the subscripts of $\Pa$. In the definition of $\Ca X$, the subscript of $\Pa$ sticks to $-X$ rather than varies with the subgroup under consideration $X'$. Because $-X\subsetneq -X'$, fixing the actions of the players in $-X$ leaves more possible strategy profiles than fixing the actions of the players in $-X'$. So we have the following fact.
\begin{fact}
If $X'\subseteq X$, then $\models \Pa_{-X} X'\rightarrow \Pa_{-X'}X'$
\end{fact}
\noindent We have explained why semantically the formula is valid. It can also be proved by using the axiom system $\textbf{LPFD}$ and the soundness result in Proposition \ref{prop:soundness}, which is much simpler.
\begin{proof}
By the axiom \textbf{(II)(e)} and the definition of $\Pa_Y X$, $\vdash_\textbf{LPFD} \Pa_{-X} X'\rightarrow \Pa_{-X'}X'$ follows immediately. Then by soundness, we get the validity.
\end{proof}
\noindent The fact points us to a critical difference between $\Pa_{-X}X'$ and $\Pa_{-X'}X'$.  As we have demonstrated in the prisoners' dilemma in Table \ref{tab:PD1}, $\bigwedge_{X'\in \mathfrak{X}(X)}\Pa_{-X'}X'\rightarrow \Pa X$ is not valid.  However, $\Pa_{-X}X'$ has the following key property.
\begin{theorem}
For all covers $\mathfrak{C}(X)$ of $X$, $\models \bigwedge_{X'\in \mathfrak{C}(X)}\Pa_{-X}X'\rightarrow \Pa X$.
\end{theorem}
\noindent Proving the validity in a model theoretical way is formidable.   By making use of the axiom system \textbf{LPFD} and its soundness, the theorem can be proved easily.
\begin{proof}
First, observe that for any $x\in X'$, $\vdash_\mathbf{LPFD} \br{=_{-X},\preceq_{X'-\set{x}}, \prec_{x}}\bot\rightarrow \br{=_{-X},\preceq_{X-\set{x}}, \prec_{x}}\bot$ follows from the axiom \textbf{(II)(e)}, because $X'-\set{x}\subseteq X - \set{x}$. Second, take an arbitrary cover $\mathfrak{C}(X)$ of $X$. Since $\bigcup_{X'\in \mathfrak{C}(X)} X' = X$, it follows immediately that $\vdash_\mathbf{LPFD} \wedge_{X'\in \mathfrak{C}(X)}\wedge_{x\in X'}\br{=_{-X},\preceq_{X'-\set{x}}, \prec_{x}}\bot\rightarrow \bigwedge_{x\in X}\br{=_{-X},\preceq_{X-\set{x}}, \prec_{x}}\bot$. By the soundness of \textbf{LPFD} and the definition of $\Pa_{-X}X'$ and $\Pa X$, it follows that  $\models \bigwedge_{X'\in \mathfrak{C}(X)}\Pa_{-X}X'\rightarrow \Pa X$.
\end{proof}
\begin{corollary}
$\models \Ca X\rightarrow \Ca_1 X$.
\end{corollary}

Therefore, the definition of $\Ca X$ ensures that the state of a group with collective agency is Pareto optimal and each agent in the group keeps their agency. In particular, our definition of collective agency addresses the issue troubling $\Ca_1 X$, namely it may happen that none of its non-singleton subgroups have reached Pareto optimality. The definition of $\Ca X$ guarantees that there is a cover of $X$ whose members all reach Pareto optimality. Readers can check that $(\text{coop,coop})\models \Ca X$ in Table \ref{tab:PD2} (hint: taking $\mathfrak{C}(\set{1,2,3}) = \set{\set{1,3},\set{2,3}$}). Compared with $\Ca_1 X$, $\Ca X$ tells us more about the structure of $X$ at $(\text{coop,coop})$. This alludes to deeper issues, for example, the role of the structure, organization, and norms of a group in shaping its collective agency.

\section{Conclusion and Future Work}\label{sec:CandFW}
We have proposed a formal approach to understanding  collective agency and demonstrated its fruitfulness. It complements the philosophical discussion of collective agency in much the same way as epistemic logic and the formal analysis of knowledge that is based on it complement epistemology. The approach also opens up a large area for further exploration, for example, modeling collective agency in extensive games with imperfect information so as to take temporal and epistemic dimensions into consideration. In particular, the temporal dimension brings out issues of the genesis, maintenance and change of collective agency. To maintain a group's collective agency, the group's structure and organization and even norms and conventions shared by the group members play a role. We will take these aspects of collective agency into consideration in our future work.


\bibliographystyle{eptcs}
\bibliography{ReferenceForLPFDSG}
\end{document}